\documentclass[11pt]{article}
\pdfoutput=1
\usepackage{graphicx}  
\usepackage{amsmath}  
\usepackage[left=1in,top=1in,right=1in,bottom=1in]{geometry}
\usepackage{authblk}   
\usepackage{setspace}  
\usepackage{cite} 
\usepackage{hyperref} 

\usepackage{mathptmx}
\usepackage[T1]{fontenc}

\newcommand{\IPL}{\ensuremath{I_\text{PL}}}
\newcommand{\fAM}{\ensuremath{f_\text{AM}}}
\newcommand{\fMW}{\ensuremath{f_\text{MW}}}
\newcommand{\muB}{\ensuremath{\mu_\text{B}}}

\begin{document}

\title{Electron spin resonance of nitrogen-vacancy centers in optically trapped nanodiamonds}
\author{Viva R.\ Horowitz}
\author{Benjam\'in J.\ Alem\'an}
\author{David J.\ Christle}
\author{Andrew N.\ Cleland}
\author{David D.\ Awschalom\footnote{To whom correspondence should be addressed. E-mail: awsch@physics.ucsb.edu}}
\affil{Center for Spintronics and Quantum Computation, University of California, Santa Barbara, California 93106, USA}
\date{}

\maketitle

\doublespace

\pdfbookmark{Abstract}{Abstract}
\noindent
\textbf{
Using an optical tweezers apparatus, we demonstrate three-dimensional control of nanodiamonds in solution with simultaneous readout of ground-state electron-spin resonance (ESR) transitions in an ensemble of diamond nitrogen-vacancy (NV) color centers.  Despite the motion and random orientation of NV centers suspended in the optical trap, we observe distinct peaks in the measured ESR spectra qualitatively similar to the same measurement in bulk. Accounting for the random dynamics, we model the ESR spectra observed in an externally applied magnetic field to enable d.c.\ magnetometry in solution. We estimate the d.c.\ magnetic field sensitivity based on variations in ESR line shapes to be $\sim$50~$\mu\text{T}/\sqrt{\text{Hz}}$. This technique may provide a pathway for spin-based magnetic, electric, and thermal sensing in fluidic environments and biophysical systems inaccessible to existing scanning probe techniques.
}

\begin{center}
{nitrogen-vacancy center $|$ nanodiamond $|$ ESR $|$ magnetometry $|$ optical tweezers}

 {Abbreviations: ESR, electron spin resonance; NV, nitrogen-vacancy}
\end{center}


\pdfbookmark{Introduction}{Introduction}

{T}he room temperature quantum coherence and optical addressability of negatively charged nitrogen-vacancy (NV) color center spins in diamond\cite{Jelezko2004} make NV centers particularly effective for a variety of sensing applications. The spin-dependent fluorescent readout of ground-state electron-spin resonance (ESR) transitions in NV centers has been used for single-spin magnetic~\cite{Degen2008, Taylor2008,   Balasubramanian2008,  Maze2008, Maertz2010, Schoenfeld2011, Pham2011, Arcizet2011,  Rondin2012,Maletinsky2012}, electric~\cite{Dolde2011}, and thermal~\cite{Toyli2012} metrology at the nanoscale. The photostability and biocompatibility of fluorescent NV centers within nanodiamonds~\cite{Yu2005, Fu2007, Liu2007, Schrand2007, Chang2008} have also permitted quantum control of NV centers within living cells~\cite{McGuinness2011}, pointing to potential applications of sensing, tracking, and tagging in submicron biophysical systems. While techniques utilizing scanning probe tips have been the focus of intensive efforts for precise spatial control of nanodiamonds~\cite{Balasubramanian2008, vanderSar2009, Barth2009, Cuche2009Near, Arcizet2011,Rondin2012, Maletinsky2012, Degen2008, Taylor2008}, this approach is less suitable within complex environments such as microfluidic channels or the interiors of biological cells.

Laser-based optical trapping is another method of precise nanopositioning, but it occurs without physical contact. Optical tweezers utilize tightly focused light to non-invasively trap and move small dielectric particles in three dimensions\cite{Neuman2004, Sun2001}. This powerful and biocompatible technique has allowed investigation of molecular motors~\cite{Block1990}, cell-sorting of a population of \textit{Eschericha coli} based on single-cell viability~\cite{Ericsson2000}, and even the observation of single-base-pair stepping ($3.7\,\text{\AA}$) of RNA polymerase along DNA\cite{Abbondanzieri2005}. With ultrastable, dual-beam optical tweezers achieving repeatable displacements at the nanometer scale and below~\cite{Abbondanzieri2005,Svoboda1993}, the prospect of combining optical tweezers with quantum-based sensors is particularly attractive for biosensing.

We demonstrate a biocompatible approach to scanning nanodiamond magnetometry in solution using a single-beam optical tweezers apparatus. The optical trap uses the radiation pressure of a focused infrared laser beam to attract and hold an ensemble of diamond nanoparticles at the focus, while a second confocal green laser optically excites the embedded NV centers. The spin-dependence  of the NV center's luminescence, together with a nearby microwave antenna, allow us to perform optically detected ESR measurements with simultaneous three-dimensional control in solution. We develop a model of the observed ESR spectra based on the ground-state Hamiltonian that accounts for the random motion of NV centers in the trap and incorporates the orientation-dependent absorption and luminescence collection efficiency. Using this model, we are able to infer the magnetic field experienced by the ensemble of NV centers and show an estimated magnetic sensitivity of $\sim\!50\,\mathrm{\mu T/\sqrt{Hz}}$.


\pdfbookmark{Results and Discussion}{Results and Discussion}
\section*{Results and Discussion}   
 \noindent  \textbf{Optically Trapping Nanodiamonds.}
We study nanodiamond ensembles in a home-built confocal microscope that combines optical trapping, NV-center optical excitation, and fluorescence detection. The optical trapping is performed with a  1064\,nm  continuous wave  laser while a separate 532\,nm  continuous wave laser is used for optical excitation of the NV centers. Both beams are tightly focused using an oil-immersion objective ($\text{NA} = 1.3$). A dichroic mirror and optical filter are used to collect the red-shifted fluorescence from the phonon side band of the NV center, while a separate notch filter is used to remove the laser scatter from the trapping laser. The filtered light is focused into a single-photon counting avalanche photodiode, whose counts are read out with a data acquisition card (See \textit{Supporting Information} for details).  All measurements are performed at room temperature.

\begin{figure}   \centering
\includegraphics{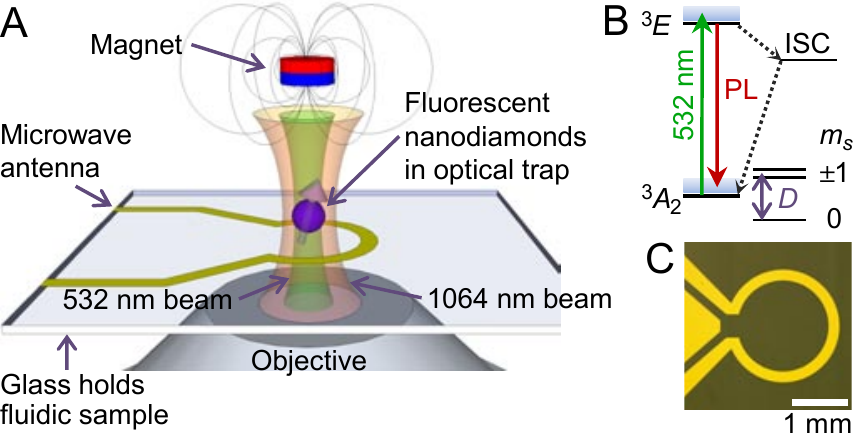}
\caption{ 
\normalsize
	(\textit{A})~Schematic of nanodiamond optical trapping, photoexcitation, and luminescence detection at the focus of the objective.   The magnetic field is applied externally, along the axis of the objective. The microwave antenna and glass coverslip are also shown. (\textit{B})~Energy level diagram of the diamond NV center. The $^{3}\!{A}_{2}$ ground state is expanded to show the spin sublevels split by the zero-field splitting, $D$. The spin system is optically excited by 532\,nm laser into the excited state ($^3\!E$), where it has a spin-dependent probability of either returning to the ground state with a red-shifted photoluminescence (PL) or decaying non-radiatively through the intersystem crossing (ISC). (\textit{C})~Micrograph of the 50-$\Omega$-impedance-matched antenna that drives coherent transitions between spin states.} 
\label{fig:CartoonAndLevelsAndAntenna}
\end{figure}

\begin{figure} \centering
\includegraphics{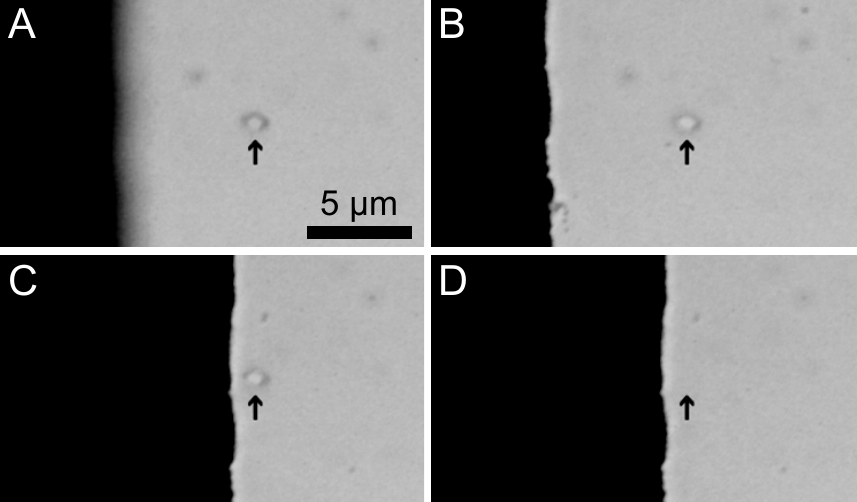}
\caption{ 
\normalsize
Spatial control of optically trapped nanodiamonds near the (black) microwave antenna (individual frames from movie, enhanced online). (\textit{A-B})~The antenna is brought into focus, moving axially by 4.2\,$\mu$m with respect to the trapped nanodiamonds from (\textit{A}) to (\textit{B}). An arrow indicates the position of the optical trap, with nanodiamonds visible in both frames. (\textit{C})~The antenna is moved laterally by 8.75\,$\mu$m while the nanodiamonds remain trapped. (\textit{D})~The trapping laser is blocked, releasing the nanodiamonds and allowing Brownian motion to scatter them away from the focus.
} \label{fig:trapsequence}
\end{figure}

The measurement geometry at the objective is shown in Fig.~\ref{fig:CartoonAndLevelsAndAntenna}\textit{A}. A drop of aqueous nanodiamond solution (See \textit{Materials and Methods} for details) is placed on a glass coverslip and brought to the focus of the objective. The relative position of the sample with respect to the beam is adjusted with an XYZ piezoelectric stage. The sequence of micrographs in Fig.~\ref{fig:trapsequence} shows an ensemble of nanodiamonds held by the optical trap near the edge of a microwave antenna patterned lithographically on the coverslip.  These images demonstrate lateral and axial control of the particles in suspension.

\vspace{.15in} \noindent  \textbf{Electron Spin Resonance Experiments with Trapped Nanodiamonds.}
The optical spin polarization and spin-dependent photoluminescence intensity, $\IPL$, of NV centers enable optically detected ESR measurements.  When combined with electromagnetically and thermally sensitive spin states,  optically detected ESR permits the measurement of the local electric, magnetic, or thermal environment. The negatively charged NV center defect in diamond consists of a substitutional nitrogen atom adjacent to a vacancy in the diamond lattice.   The NV center's unperturbed electronic energy level structure, shown in Fig.~\ref{fig:CartoonAndLevelsAndAntenna}\textit{B}, consists of a ground-state spin triplet with lowest-energy spin sublevel $m_s = 0$ along with two  $m_s = \pm 1$ spin sublevels~\cite{Gruber1997,Manson2006}, which are nominally degenerate at zero magnetic field and energetically higher than $m_s=0$ by the crystal field splitting, $D=2.87$\,GHz.  The energy of the NV center spin system is magnetically sensitive, much like that of its classical analogue, the magnetic dipole.  Specifically, a magnetic field $\mathbf{B}$ will shift the energy of the NV center's spin states according to the ground-state Hamiltonian, 
\begin{equation}\label{eq:Hamiltonian}
\hat{H}_\text{NV} = D \hat{S}_z^2 + g \muB \mathbf{B} \cdot \hat{\mathbf{S}},
\end{equation}
where $g = 2$ is the electronic $g$-factor, $\muB$ is the Bohr magneton, and $\hat{\mathbf{S}}$ is  the electronic spin operator.  The measurement of  spin energy eigenvalues in the presence of a magnetic field is the experimental basis for magnetic sensing using NV centers.  The optical read-out of the spin state is possible because the $m_s = \pm 1$ states have a higher probability of a non-radiative transition via the inter-system crossing (ISC), so $\IPL$ is lower in these states than in the brighter $m_s = 0$ state.  Control of the spin state is achieved with a combination of optical and microwave pumping:  optical excitation initializes the system into the $m_s = 0$ state through the same ISC, while a microwave field resonant with the energy splitting between the $m_s = 0$ and the $m_s={+1}$ or ${-1}$ states will coherently rotate the spin into a superposition of the spin sublevels, which we detect as a darker $\IPL$.  In order to apply microwave fields, we designed a microwave antenna that is lithographically patterned on the glass coverslip and impedance-matched near $2.87\,\text{GHz}$ to optimize power transmission and reduce heating, shown in Fig.~\ref{fig:CartoonAndLevelsAndAntenna}\textit{C}.  In continuous wave ESR measurements, the photoluminescence intensity ($\IPL$) is read out under the continuous application of both the 532\,nm laser and microwave fields, leading to resonances in the observed intensity as the applied microwave field is swept in frequency across the spin sublevel transitions.  The photoluminescence contrast measured between spin states  can exceed 20\% in fixed nanodiamonds (See \textit{Supporting Information}). 

\begin{figure} \centering
\includegraphics{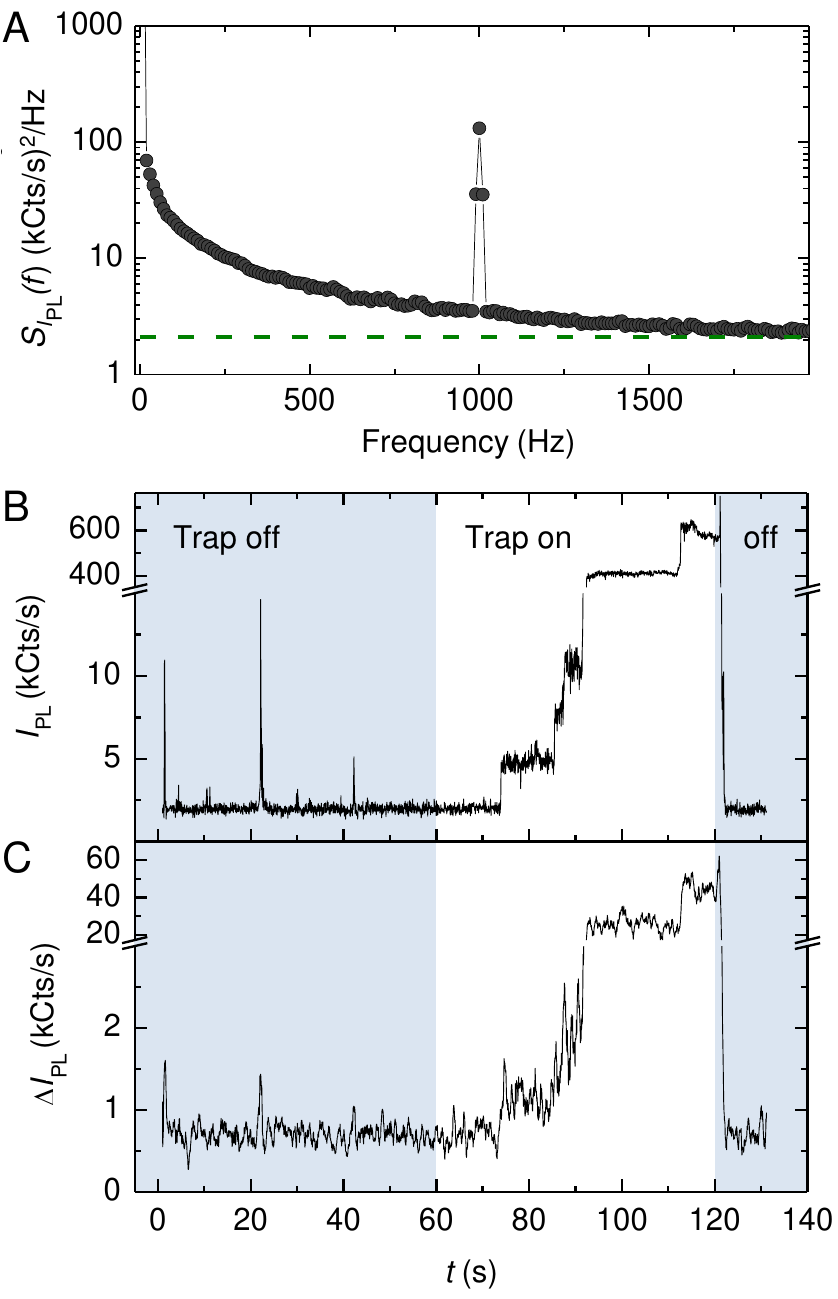}
\caption{
\normalsize
(\textit{A})~Bimodal power spectral density $S_{\IPL}(f)$ of the luminescence with on-resonant microwaves at frequency $\fMW = 2.87$\,GHz and amplitude-modulation frequency $\fAM = 1$\,kHz. The peak at 1\,kHz corresponds to the NV response to the amplitude-modulated carrier signal. The green dashed line shows the expected shot noise floor for the bimodal power spectral density, $S_{\IPL}(f) = \IPL = 2.1$\,MHz. The noise at $\fAM=1$\,kHz is higher than the expected floor, indicating that the measurement is not shot-noise limited. The power spectral density was calculated from $\IPL$ data taken for $100\,\text{ms}$, our typical lock-in time; the average of 1,000 sets of data is shown. (\textit{B})~Time trace of $\IPL$ showing discrete steps of increasing photoluminescence as clusters of NV centers enter the optical trap with the green excitation laser on. The trapping laser is initially blocked (blue shaded times). The trapping laser is unblocked at time $t = 60\,\text{s}$, and $\IPL$ remains low, indicating an empty trap, until the first discrete step at $t \sim 75$\,s. At time $t \sim 120$\,s, the trapping laser is blocked to release the particles from the trap, with $\IPL$ dropping commensurately. (\textit{C})~The coincident ESR response of resonant microwaves, applied at $\fMW = 2.87$\,GHz, indicates that the fluorescent particles in the trap are indeed nanodiamonds that contain NV centers.  Data in (\textit{C}) is smoothed.
} \label{fig:PSDandSteps}
\end{figure}

Random fluctuations in the photoluminescence of optically trapped fluorescent nanodiamonds present experimental challenges in measuring the ESR contrast.  The Brownian motion in solution, collisions between nanoparticles, and the entry and exit of nanodiamonds from the optical trap contribute to a large, low-frequency noise component in the observed $\IPL$.  ``Blinking,'' attributed to charge instabilities related to surface effects\cite{Bradac2010}, may augment the observed fluorescence fluctuations.  To increase the signal-to-noise ratio during ESR measurements, we use commercial nanodiamonds that have been $\text{He}^{+}$ irradiated to create vacancies and subsequently annealed to form approximately 500 NV centers per $\sim$100\,nm diameter nanodiamond.  Additionally, by performing amplitude modulation of the applied microwaves with a software-based photon-counting lock-in technique\cite{Braun2002,Arecchi1966}, we improve the signal-to-noise ratio of the experiment by more than a factor of ten (See \textit{Supporting Information} for further details).  In this way, ESR dips in $\IPL$ are converted to peaks in the differential luminescence $\Delta \IPL$.  Figure~\ref{fig:PSDandSteps}\textit{A} shows the power spectral density of $\IPL$ for trapped nanodiamonds, displaying both the low-frequency noise and the NV ESR contrast response from resonant microwaves that are amplitude modulated at 1\,kHz. Figures~\ref{fig:PSDandSteps}\textit{B-C} show the measured $\IPL$ and contrast $\Delta \IPL$ before and after turning on the trapping beam. As the trapping beam remains on, fluorescent nanodiamonds stochastically enter the trap and cause $\IPL$ to increase in discrete steps, with coincident rises in $\Delta \IPL$ indicating the presence of NV centers.  When the trapping beam is turned off, the nanodiamonds scatter out of the trap from Brownian motion, causing the luminescence to cease. We typically observe the contrast $\Delta \IPL/\IPL$ at zero field to be $\sim$10\% at most; this is smaller than the contrast observed in bulk, which may be a consequence, in part, of non-NV background fluorescence in the nanoparticles. Ongoing research in the production of high purity nanodiamonds has the potential to significantly reduce these complications.

\begin{figure*}
\includegraphics{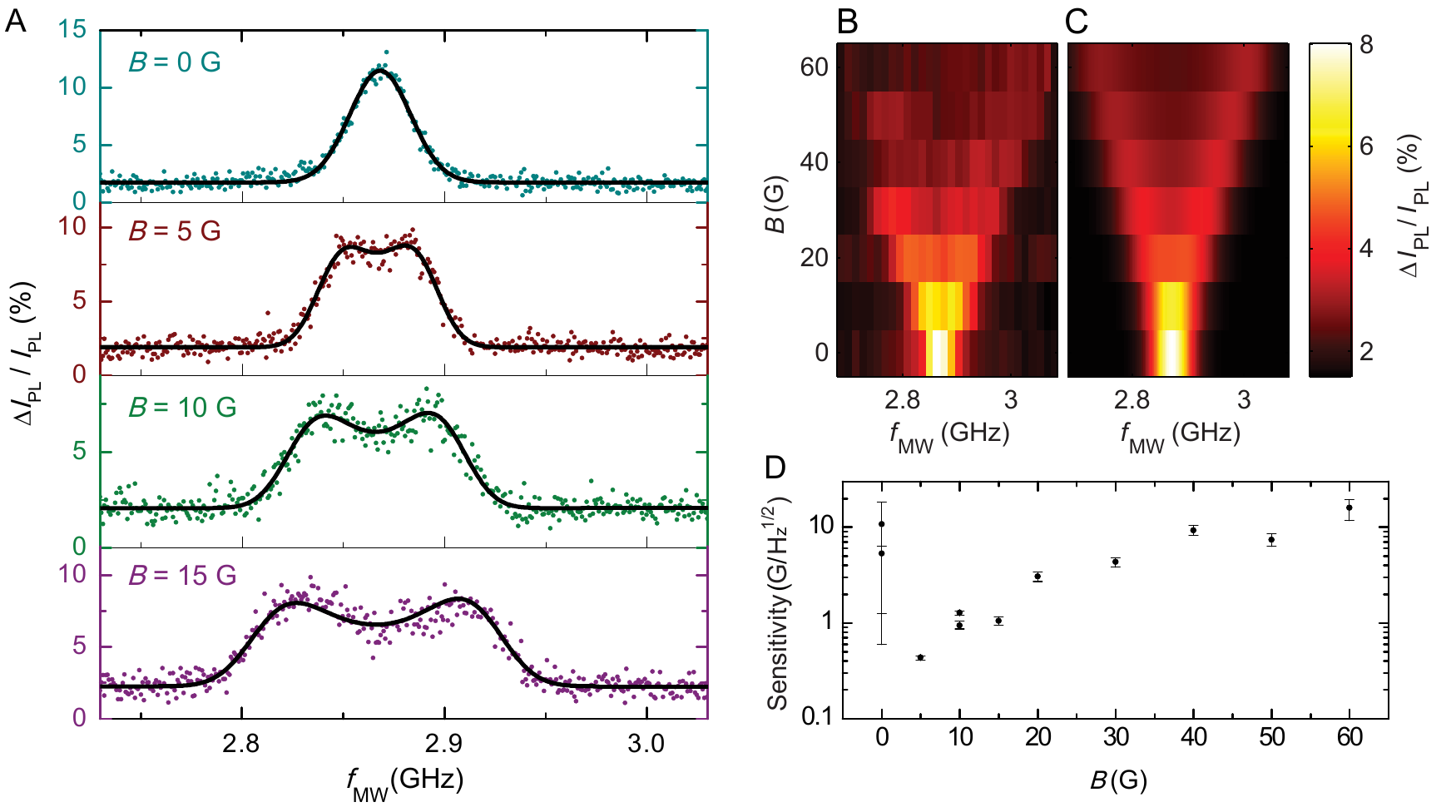}
\caption{
\normalsize
(\textit{A}) Optically detected ESR spectra of trapped nanodiamond ensembles at calibrated low field strengths. The best-fit curve from the model with five fitting parameters is shown (black lines). The measurements occurred at calibrated applied magnetic fields of 0, 5, 10, and 15\,G. By fitting to the model, we obtain estimated magnetic fields of 1.5, 9.4, 14.5 and 20.8\,G, respectively. This discrepancy is discussed further in the \textit{Supporting Information}. Each ESR spectrum has a total acquisition time of about 200\,s.  $\IPL \approx 510$\,kCts/s and $\fAM = 1$\,kHz.
(\textit{B})~Measured ESR spectra of trapped nanodiamonds up to 60\,G.
(\textit{C})~Predicted ESR spectra from the model are computed by fixing all parameters except $B$ to their best-fit values at zero field and adjusting $B$ to the calibrated field values.
(\textit{D})~Estimated sensitivity of the diamond-based magnetometer using the optimal measuring scheme. The estimates are computed using values inferred from experiment, and the error bars reflect the 68.2\% highest probability intervals from the propagated uncertainties. 
} \label{fig:ESRspectra}
\end{figure*}

Although the measured ESR signal is effectively an average over ensembles of randomly oriented, mobile NV centers, ESR spectra (Fig.~\ref{fig:ESRspectra}\textit{A}) at low magnetic fields retain qualitative similarities to measurements of aligned NV centers in bulk diamond:  the spectra exhibit two distinct peaks that shift approximately linearly ($\sim$2.5\,MHz/G) as a magnetic field is applied.  However, unlike aligned NVs, the spectral peaks broaden with increased magnetic field.  To understand the lineshape and magnetic field dependence of the observed spectra, we develop a model consisting of a statistical average over all possible NV center orientations with respect to a fixed magnetic field, incorporating the directional dependence of the transition frequencies from Eq. \eqref{eq:Hamiltonian} and the anisotropic excitation and collection efficiencies of our confocal microscope.  In this model, the orientation of the NV center's symmetry axis, relative to the magnetic field and optical axis, determines its ESR resonance frequency and contribution to the overall spectrum: a perpendicular orientation yields a minimal contrast contribution and frequency shift, while a parallel NV center gives a maximal contribution and shift (2.80~MHz/G).  Summing over an isotropic distribution of NV center orientations, we expect the overall ESR spectrum to have two broadened peaks, in accord with our experimental observations.  To infer the model parameters and their associated uncertainties for each measured spectrum, we apply a Bayesian Markov Chain Monte Carlo approach~\cite{Laloy2012} and plot the best-fit curves over the plotted data. Measured ESR spectra up to 60\,G are shown in Fig.~\ref{fig:ESRspectra}\textit{B}. In Fig.~\ref{fig:ESRspectra}\textit{C}, we fix the model parameters found at 0\,G, adjust only the parameter for the applied magnetic field, and obtain excellent qualitative agreement with the data. 

From the model, we can gain an intuition for the optimal conditions for d.c.\ magnetometry in our system.  Since the splitting of the peaks is approximately linear with increasing field  and the contrast diminishes, we would expect that the magnetic sensitivity generally worsens at higher fields.  Similarly, when the two resonances become unresolved near zero magnetic field, the line shape becomes weakly dependent on $B$ and the sensitivity is poor.  For this reason, an optimum condition exists at low fields $\sim$5\,G when the resonances are split but the contrast is still large.  We estimate from the statistical analysis the optimal magnetic sensitivities in Fig.~\ref{fig:ESRspectra}\textit{D}, which depend on the noise and lineshape inferred from experiment (See the \textit{Materials and Methods} for details).  The most sensitive estimate of $\sim$50\,$\mathrm{\mu T/\sqrt{Hz}}$, measured with $\IPL \approx 510$\,kCts/s,  occurs at low fields (${\sim}$5\,{G}) when the peaks are at least partially split.  Further improvements to the collection efficiency and operating at a higher modulation frequency could improve the sensitivity of this technique by a factor of $\sim$20, making it competitive with existing NV scanning-probe d.c.\ magnetometry protocols\cite{Schoenfeld2011, Rondin2012, Maletinsky2012}. Stable trapping of single nanodiamonds would ameliorate noise from collisions within the trap and improve the spatial resolution of the technique. The use of shaped diamond particles trapped with a controlled orientation\cite{Shelton2005} and aligned along the appropriate crystallographic axis would remove the degrees of freedom that complicate the ESR lineshape from the situation in bulk, opening the possibility of improved contrast and even vector magnetometry\cite{Maertz2010}.

\pdfbookmark{Conclusions and Outlook}{Conclusions and Outlook}
\section*{Conclusions and Outlook} 
The combination of optical trapping and NV-center-based sensing developed in this work enables the three-dimensional mapping of magnetic fields in solution and addresses the need to probe complex environments, such as the interiors of microfluidic channels.  Together, these two powerful techniques could pave the way for exploiting the unique electromagnetic and thermal sensing properties of NV centers at the nanoscale. Using optically trapped nanodiamonds for intracellular sensing~\cite{McGuinness2011}, the mapping of electrical fields and thermal gradients around cells~\cite{Duan2012, Brites2012}, or the mapping of neurons~\cite{Hall2012, Pham2011} are particularly exciting applications of this technique.  The three-dimensional position control and on-demand release of optically trapped nanodiamonds achieved herein enables applications requiring nanoscale precision placement of NV centers within existing systems, such as the controlled tagging of a single biological cell.  Additionally, this technique may serve as a tool for monitoring physical and chemical processes at liquid/solid interfaces, which could help improve the understanding of electrochemical cells, surface catalysis, or lipid membranes in biomedicine.

 \pdfbookmark{Materials and Methods}{Materials and Methods}
\section*{Materials and Methods}  
\pdfbookmark[1]{Microwave antenna microfabrication}{Microwave antenna microfabrication}
 \noindent  \textbf{Microwave antenna microfabrication.}
For antenna fabrication, 10\,nm of titanium and 1000\,nm of gold were evaporated onto the freshly piranha-etched 150\,$\mu$m thick glass coverslips 35\,mm$\,\times \,$50\,mm in dimension.  Standard photolithography was used to define a resist etch mask, then gold and titanium etchants were used to transfer the desired antenna pattern to the substrate.  Antennas were wire bonded to a coplanar waveguide on a printed circuit board, as shown in the \textit{Supporting Information}. The circuit board was then fastened down to the XYZ piezoelectric stage and connected to a microwave signal generator and amplifier for measurements.

\vspace{.15in} 
\pdfbookmark[1]{Nanodiamonds}{Nanodiamonds}
\noindent  \textbf{Nanodiamonds.}
We used commercially available synthetic HPHT type Ib nanodiamonds from Adamas Nanotechnologies.  The nanodiamonds have been {He}$^{+}$ irradiated, annealed, and purified with acids by the manufacturer. The particles, typically 100\,nm across, are specified to contain 500 or more NV centers per particle on average.  All measurements were taken in filtered, deionized water.  See the \textit{Supporting Information} for electron microscopy images of the nanodiamonds.

\vspace{.15in} 
\pdfbookmark[1]{Estimating trap population of NV centers}{Estimating trap population of NV centers}
\noindent  \textbf{Estimating trap population of NV centers.}
The number of NV centers in the trap and in the measurement volume during our experiments can be estimated from the NV density of the particles.  The density is specified to be $\sim$500 NV centers per 100\,nm diameter diamond, corresponding to a nearest neighbor separation of 12.6\,nm. We estimate the trapping and measurement volume from the beam waist ($w_0={n \lambda}/{\pi \text{NA}}$) and Rayleigh range ($\pi {w_0^2}/{\lambda}$) of a focused Gaussian beam. The volume is given by the expression $V=\frac{n^4\lambda^3 }{4 \pi^2 \,\text{NA}^4}$, where $n$ is the index of refraction ($n=1.515$), $\text{NA}$ is the numerical aperture of the objective ($\text{NA} =1.3$), and $\lambda$ is the laser wavelength.  Using this approximation, the trapping volume is $V_\text{trap}=0.06\,\mathrm{\mu m^3}$ and the measurement volume is $V_\text{msr}=0.007\,\mathrm{\mu m^3}$.  The trap width is $w_0\approx 0.4\,\mathrm{\mu m}$, in good agreement with the width observed in optical images (see Fig. \ref{fig:trapsequence}.)  Assuming a unity packing-fraction, the maximum number of 100\,{nm} diameter particles, each of approximate volume $0.001\,\mathrm{\mu m^3}$, that can occupy the trap and measurement volumes is 60 and 7, respectively.  In terms of NV centers, the upper bound in the measurement volume is approximately 3,500 centers.  The highest stable $\IPL$ we observed in our experiments was 3,000\,{kCts/s}.  Assuming this value corresponds to a full trap with $\sim$3,500 NV centers, we would expect each NV to contribute $\sim$0.86\,{kCts/s}.  This value agrees with the experimentally measured value of 1\,kCts/s, obtained by measuring the minimum step height of $\IPL$ as 35\,nm diameter nanodiamonds (also from Adamas Nanotechnologies), each specified to contain approximately 1 to 4 NV centers, enter the optical trap.

\vspace{.15in}
\pdfbookmark[1]{Modeling ESR spectra}{Modeling ESR spectra}
 \noindent  \textbf{Modeling ESR spectra.}
We model the magnetic field dependence of the ESR signal by assuming the measurement volume contains a large ensemble of isotropically oriented NV centers. The large ensemble assumption is justified by an estimate for the NV population in the trap's measurement volume (see above) that yields approximately 3,500 NV centers. The orientation of an NV center with respect to the magnetic field vector determines the level splitting according to the Hamiltonian in Eq.~\eqref{eq:Hamiltonian}, while the orientation with respect to the objective affects the strength of the optical absorption and the collection efficiency of the total emitted photoluminescence. From geometrical considerations of the two NV center transition dipoles~\cite{Epstein2005}, we approximate the angular dependence of the absorption of each NV center to be $1+\cos^2 \theta$, where $\theta$ is the angle between the NV symmetry axis and the magnetic field vector. We integrate the far-field emission of each transition dipole over the collection cone of a 1.3 numerical aperture oil-immersion objective to account for the angular dependence of the collected luminescence signal. To obtain the final spectral shape, the individual splittings for each orientation are convolved with a Gaussian function to account for the natural linewidth with power broadening and finally integrated over an isotropic orientation density.

Since the typical ESR contrast in experiment is around 7\% and the measurements are not shot-noise limited (by a factor of approximately two), the error at each point is assumed to be identical (homoscedastic) and normal, and is treated as a free parameter in the model. Further technical details regarding the analysis are found in the \textit{Supporting Information}.

\vspace{.15in} 
\pdfbookmark[1]{Magnetic sensitivity calculations}{Magnetic sensitivity calculations}
\noindent  \textbf{Magnetic sensitivity calculations.}
The theoretical magnetic sensitivity is related to the noise and the lineshape of the associated ESR spectrum. At a given microwave frequency, $\fMW$, small changes in the measured contrast signal $C \equiv \Delta \IPL / \IPL $ occur with a change in magnetic field according to $\delta C = \delta B \, \frac{\partial C}{\partial B}$ and thus the most efficient magnetometry measurement would take place at the $\fMW$ where this derivative is largest in magnitude. In the low-field limit, this maximum in $\left|\frac{\partial C}{\partial B}\right|$ occurs for $\fMW$ centered between the two peaks, approximately $\fMW = 2.87$\,GHz; however, once the two peaks split by about twice the FWHM, the most sensitive $\fMW$ for measurement occurs on the downward slope of the highest frequency peak.  If the minimum detectable change in magnetic field is $\delta B_\text{min}$, then the estimated optimal magnetic sensitivity is~\cite{Taylor2008, Dreau2011}
\begin{equation}
	\eta_B = \delta B_\text{min}  \sqrt{\Delta t} =
		\frac{\sigma_C \, \sqrt{\Delta t}} {\text{max} \left| \frac{\partial C}{\partial B} \right|} ,
\end{equation}
where $ \sigma_C$ is the estimated standard deviation of $C$ for measurement time $\Delta t$ from the analysis.

We calculate the maximum-magnitude derivative of the model with respect to the parameter $B$ and corresponding error parameter over samples from the Markov Chain Monte Carlo output to obtain a probability density for the optimal sensitivity given the lineshape inferred from experiment. The mean estimated and 68.2\% highest probability density intervals from this distribution are plotted in Fig.~\ref{fig:ESRspectra}\textit{D}~\cite{ChenShao1999}.
In addition, the standard deviation of the marginal density for $B$, scaled by the square root of the total acquisition time, serves as an empirical measure of the actual sensitivity obtained in the experiment, and is plotted in the \textit{Supporting Information} for comparison to the optimal estimates.

\pdfbookmark{Acknowledgments}{Acknowledgments}
\section*{\large Acknowledgments}    
We thank Lee C.\ Bassett for discussion and assistance with figure preparation, Jayna B.\ Jones and Daniel Kirby for initial work on the apparatus, and Paolo Andrich for technical assistance. We thank Stephan Kraemer of the  Microscopy and Microanalysis Facility at the UCSB Materials Research Laboratory  for assistance with the TEM.  We acknowledge the support of CNSI. A portion of this work was done in the UCSB nanofabrication facility, part of the NSF-funded NNIN network. We acknowledge the support of DARPA and AFOSR.  B.J.A.\ acknowledges support from the University of California President's Postdoctoral Fellowship.

\section*{\large Competing Financial Interests statement}
The authors declare that they have no competing financial interests. 

\section*{\large Author Contributions}
V.R.H., B.J.A., A.N.C, and D.D.A designed the experiment;  V.R.H. and B.J.A. performed the experiment;  V.R.H., B.J.A., and D.J.C. worked on fabrication and analyzed the data;  V.R.H. built the optical apparatus; V.R.H. and B.J.A. constructed the ESR measurement platform; all authors contributed to writing the paper.

\pdfbookmark{References}{References}

\newpage

\pdfbookmark{Supporting Information}{Supporting Information}

\renewcommand{\thefigure}{S\arabic{figure}}
\renewcommand{\thetable}{S\arabic{table}}
\renewcommand{\theequation}{S\arabic{equation}}
\renewcommand{\thepage}{S\arabic{page}}  

\noindent
{\LARGE \textbf{Supporting Information} }
\appendix
%
\vspace{-.1in}
\section{Nanodiamond samples}
 
\begin{figure*}  \centering 
\includegraphics{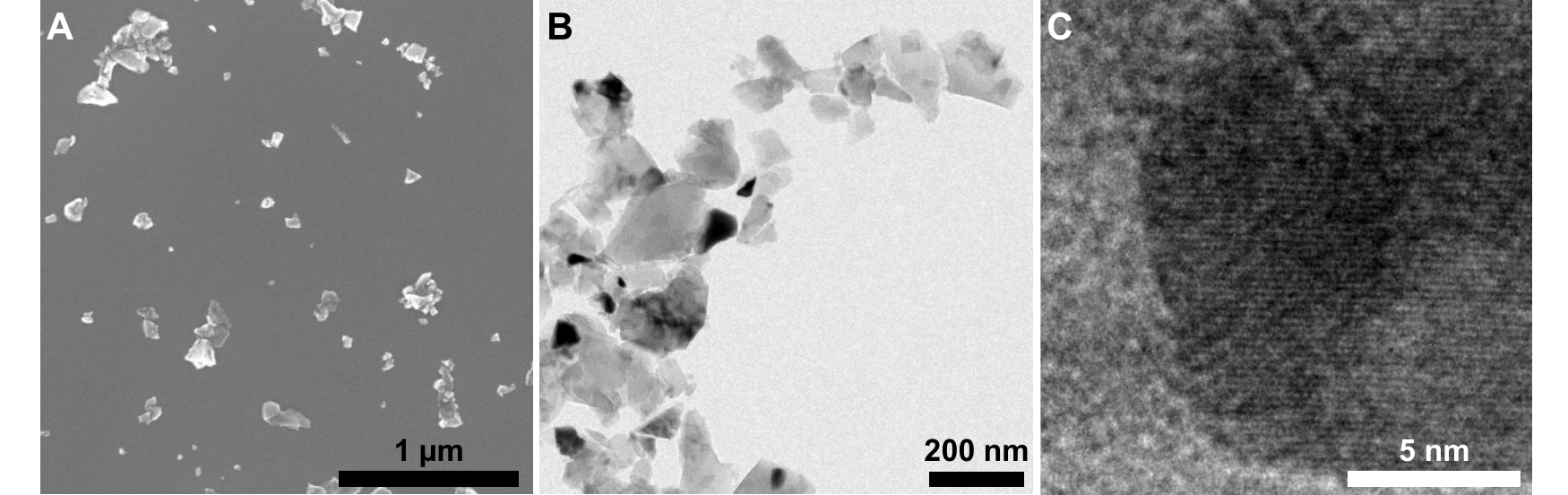}
\caption{ 
	(\textit{A})~Scanning electron micrograph and (\textit{B-C})~transmission electron micrograph of nanodiamonds  (ND-500NV-100nm, Adamas Nanotechnologies).
}\label{fig:EMofVFND}
\end{figure*}

 Figure~\ref{fig:EMofVFND} shows electron microscopy images of the nanodiamonds. These particles are synthetic HPHT type Ib diamond, approximately 100~nm in size, and irradiated with $\text{He}^{+}$ ions to create vacancies and annealed. The resulting product is purified using acids. The nanodiamonds are specified by the manufacturer (Adamas Nanotechnologies) to contain on average $>$500 NV centers/particle. 
We also used $0-0.2 \, \mathrm{\mu m}$ nanodiamonds from Microdiamant that are not irradiated or annealed in some of the measurements in this \textit{Supporting Information}.

\section{Fluorescence spectrum}

\begin{figure}  \centering
\includegraphics{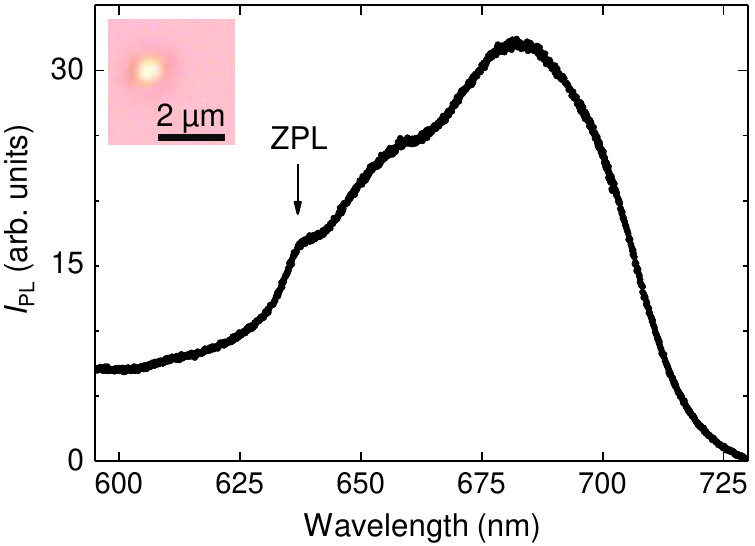}
\caption{ 
Photoluminescence spectrum of an ensemble of optically trapped nanodiamonds.  The arrow marks the NV zero phonon line.  The spectrum of the phonon sideband is attenuated for wavelengths longer than 700~nm due to a dichroic optical filter that reflects the trapping laser into the objective.  These nanodiamonds are not irradiated.
The photographic inset shows this nanodiamond ensemble in the optical trap.
} \label{fig:Spectrum}
\end{figure}

A fluorescence spectrum of optically trapped nanodiamonds is shown in Fig.~\ref{fig:Spectrum}.  A significant portion of the signal is lost due to the optical tweezers dichroic filter.  One improvement to the collection efficiency would be to replace this dichroic with one that transmits rather than reflects wavelengths from 700 nm to 800 nm.

\begin{figure*}  \centering
\includegraphics{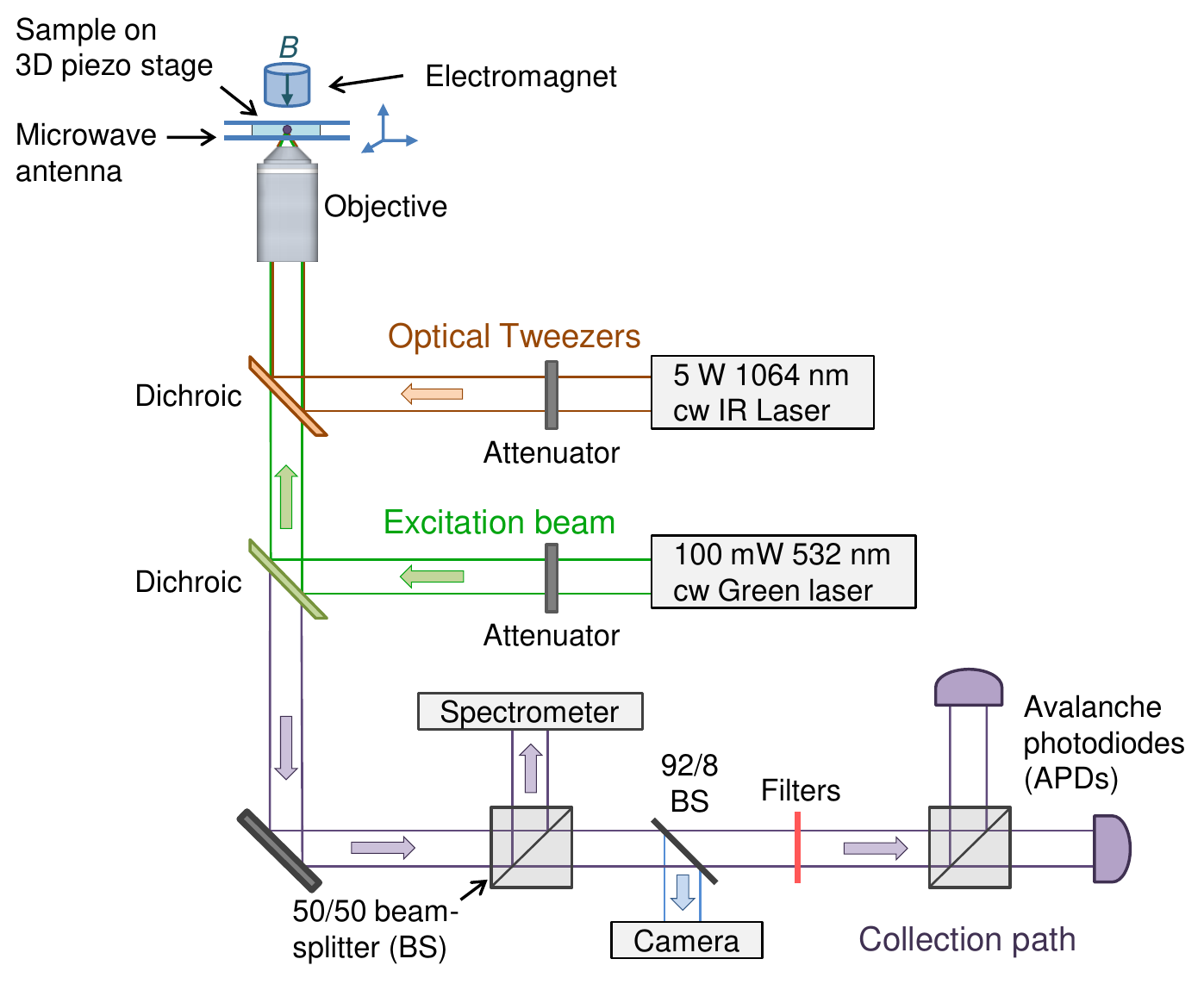}
\caption{ 
	 The apparatus, with an optical tweezers path, an excitation beam path, and a collection path.  All three optical paths are adjusted to the sample focus at the sample so that the photoluminescence signal is collected from the same confocal region where the nanodiamonds are trapped.  During measurements, the trapping location remains fixed while a 3-axis piezoelectric stage controls the sample position.
} \label{fig:apparatus}
\end{figure*}

\section{Apparatus and techniques}

 Figure~\ref{fig:apparatus} shows a schematic of the optical apparatus.   A 5~W continuous wave 1064~nm laser (NP Photonics seed laser and PM-ASA-SFA-5W amplifier, Nufern) optically traps nanodiamonds in solution in water.  A 100~mW continuous wave 532~nm laser (GCL-532-100-L CW DPSS, CrystaLaser) excites photoluminescence and polarizes the NV spin into the \(m_s=0\) spin state.  These lasers are attenuated to 30~mW and 90~$\mu$W, respectively, measured at the back opening of the objective. The photoluminescence signal is detected by an avalanche photodiode single photon counting module (APD, SPCM-AQRH-13-FC, Perkin Elmer) whose pulses are counted by a DAQ (National Instruments).  The 1064~nm and 532~nm beams are combined using dichroic mirrors (Chroma) mounted in a pair of dichroic turrets built into an inverted microscope (TE2000U, Nikon). An oil-immersion objective (CFI Plan Fluor 100x, Nikon) with numerical aperture 1.3 focuses the beams onto the sample.  We mount the sample on a 3-axis piezoelectric stage (P-517.3CL with  E-710.4CL controller, Physik Instrumente), which enables moving the antenna/coverslip assembly by up to 100 microns in X and Y and up to 20 microns in Z.  We calibrate the electromagnet (EM050-6H-222, APW Company) with a Hall probe (HMMA-1808-VR probe and 455 DSP Gaussmeter, Lakeshore).  A Hewlett Packard ESG-D4000A generates the microwave signal, which is amplified by an Amplifier Research 5S1G4. The fluorescence spectrum is measured with a SpectraPro 2750 spectrometer (PI Acton).   A 92/8 pellicle beamsplitter directs a fraction of the optical signal to a CMOS color camera (PixeLINK).  The optical signal is filtered  with a  640~nm long pass filter and a 1064 nm notch block filter to remove laser scatter prior to APD photon detection.  All optical measurements were taken with room lights off to avoid extra photon counts.

\begin{figure*}  \centering
\includegraphics{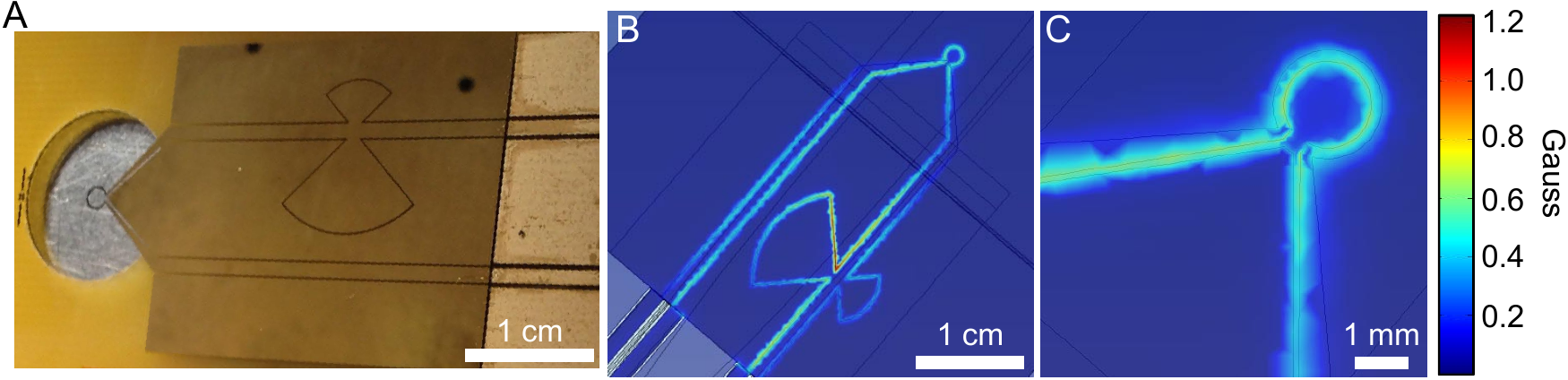}
\caption{ 
(\textit{A})~Photograph of the antenna/coverslip assembly.  The hole in the antenna mount under the antenna loop permits optical access.
(\textit{B-C})~Magnetic flux density norm in the plane of the CPW antenna when $\fMW = 2.8$~GHz, modeled in COMSOL Multiphysics.  Irregularities in the simulated magnetic flux density norm, appearing as splotching near the antenna trace edges, are an artifact of the chosen finite element meshing.
} \label{fig:Antenna_Simulation}
\end{figure*}

The antenna/coverslip assembly is shown in Fig.~\ref{fig:Antenna_Simulation}\textit{A}.  The antenna is impedance-matched to 50~$\Omega$.  The design was developed using COMSOL Multiphysics simulations.  The magnetic flux density in the vicinity of the antenna resulting from a microwave field is shown in Fig.~\ref{fig:Antenna_Simulation}\textit{B-C}.

\begin{figure}  \centering 
\includegraphics{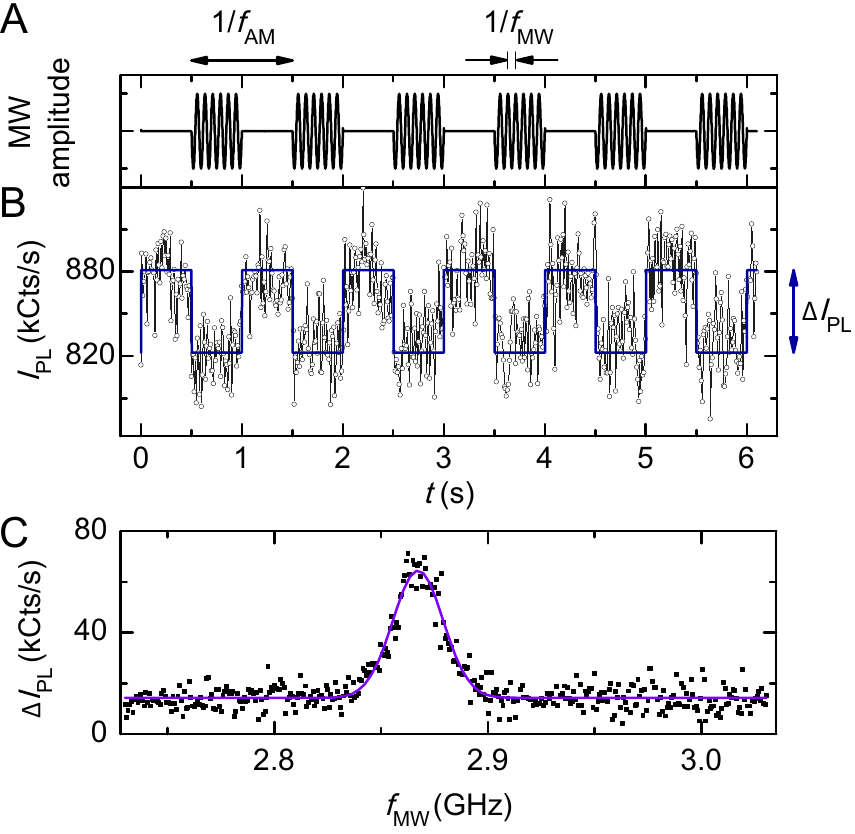}
\caption{ 
	Amplitude-modulated ESR of NV centers in optically trapped nanodiamonds in water. These nanodiamonds are irradiated.  (\textit{A})~Schematic of the amplitude-modulated microwave signal used in the experiments, with  $\fMW$ lowered for illustration.  (\textit{B})~The readout contrast of the fluorescence signal created by the modulation amplitude of resonant,  $\fMW = 2.868$~GHz, microwaves, modulated at frequency $\fAM = 1$~Hz.  We use a software lock-in (blue line) to extract the differential fluorescence intensity $\Delta \IPL = 58.5$ kCts/s or the relative ESR signal $\Delta \IPL/ \IPL = 6.64\%$.
(\textit{C})~Optically detected ESR spectrum obtained by sweeping $\fMW$ while $\fAM = 1$~kHz.  The Gaussian fit (purple line) has a FWHM of 27.8 MHz and a peak at 2.87~GHz, which is the zero-field splitting between the $m_s={0}$ and the $m_s={\pm{}1}$ levels.  This ESR spectrum was collected in 150~s. The bandwidth of the software lock-in is approximately 1 to 10~Hz.
} \label{fig:SquareOscillationsAndESR}
\end{figure}

The software lock-in is shown in Fig.~\ref{fig:SquareOscillationsAndESR}\textit{A-B}.  Since $\fMW$ is resonant with the energy splitting between the $m_s=0$ and the $m_s={+1}$ or ${-1}$ states,   $\IPL$ drops while the microwave is on, such that $\IPL$ oscillates in time at frequency $\fAM$.  Locking in to the signal, we extract the differential fluorescence signal $\Delta \IPL$.    As we sweep the microwave frequency $\fMW$, $\Delta \IPL$ remains low when $\fMW$ is off resonance with the transition between spin states and increases when $\fMW$ is on resonance. If $X$ and $Y$ are the two output channels of the lock-in, and $R = \sqrt{X^2+Y^2}$, then $\Delta \IPL = 2R$.  Figure~\ref{fig:SquareOscillationsAndESR}\textit{C} shows the ESR spectrum for an ensemble of trapped nanodiamonds in water with no externally applied magnetic field.  The nanodiamonds are specified to be 100~nm in diameter and contain 500 NV centers each.  The spectrum has a linewidth of 23.6~MHz  and a maximum  at 2.87~GHz, agreeing with the expected zero-field splitting of the NV center.   Off resonance, the curve does not go to zero because the lock-in is not phase locked.

\section{$\IPL$-dependence of noise}

\begin{figure}  \centering
\includegraphics{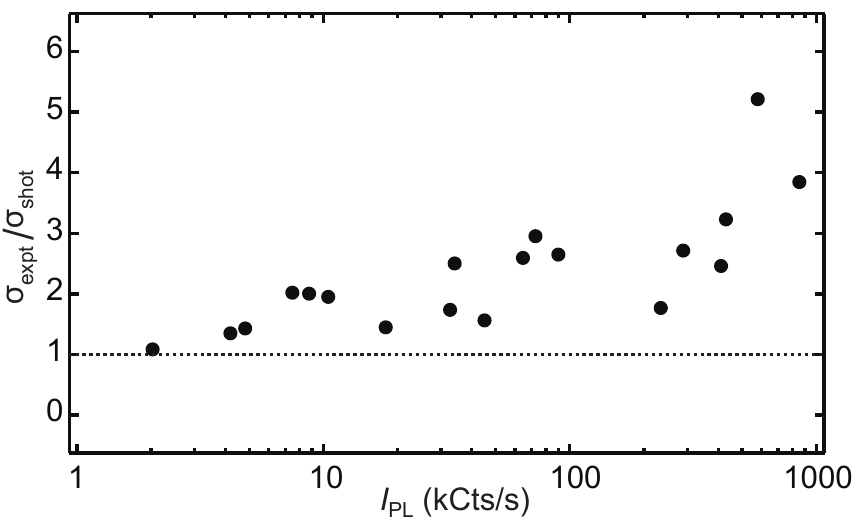}
\caption{
      The ratio of the measured standard deviation of $\IPL$ to shot noise obtained from total counts in a time interval $\Delta t$, $\sigma_\text{shot}=\sqrt{\IPL\Delta t}$, as a function of $\IPL$.  The dotted line corresponds to the case when the measured noise equals the shot noise.  This  illustrates the increase in overall photoluminescence noise observed as the optical trap becomes more populated with fluorescent nanodiamonds.
} \label{fig:Shot_Noise_vs_Measured_Noise}
\end{figure}

The photoluminescence noise from optically trapped fluorescent nanodiamonds shows a dependence on the photoluminescence, $\IPL$.  In general, the standard deviation of the experimentally measured $\IPL$, $\sigma_\text{expt}$,  grows with increased $\IPL$ beyond that expected from Poisson statistics or shot noise behavior, namely $\sigma_\text{expt}>\sigma_\text{shot}=\sqrt{N}$, where $N = \IPL \Delta t$, and $\Delta t$ is the time interval in which photon counts are measured.  Figure~\ref{fig:Shot_Noise_vs_Measured_Noise} illustrates this dependence and plots the ratio $\sigma_\text{expt}/\sqrt{N}$ as a function of $\IPL$.  For low values of $\IPL$, $\sigma_\text{expt}$ approaches shot noise (dotted line in Fig.~\ref{fig:Shot_Noise_vs_Measured_Noise}), but reaches values nearly 6 times shot noise at higher $\IPL$.  Analysis of noise is performed only on plateaus with stable $\IPL$, therefore the plotted noise, especially for higher $\IPL$, is a lower limit.  As discussed in the text, this increase in noise likely arises from dynamics of multiple particles in the trap. Any motion of particles in the trap, arising from collisions, thermal vibrations, trap instability, \textit{etc.}, will have a corresponding contribution to the noise in $\IPL$ because each particle will experience a varying degree of electric field strength from the laser beam determined by its position in the measurement volume.  
In addition, when a fluorescent particle enters the trap, the abrupt increase in $\IPL$ contributes to the noise.

\begin{figure*}  \centering
\includegraphics{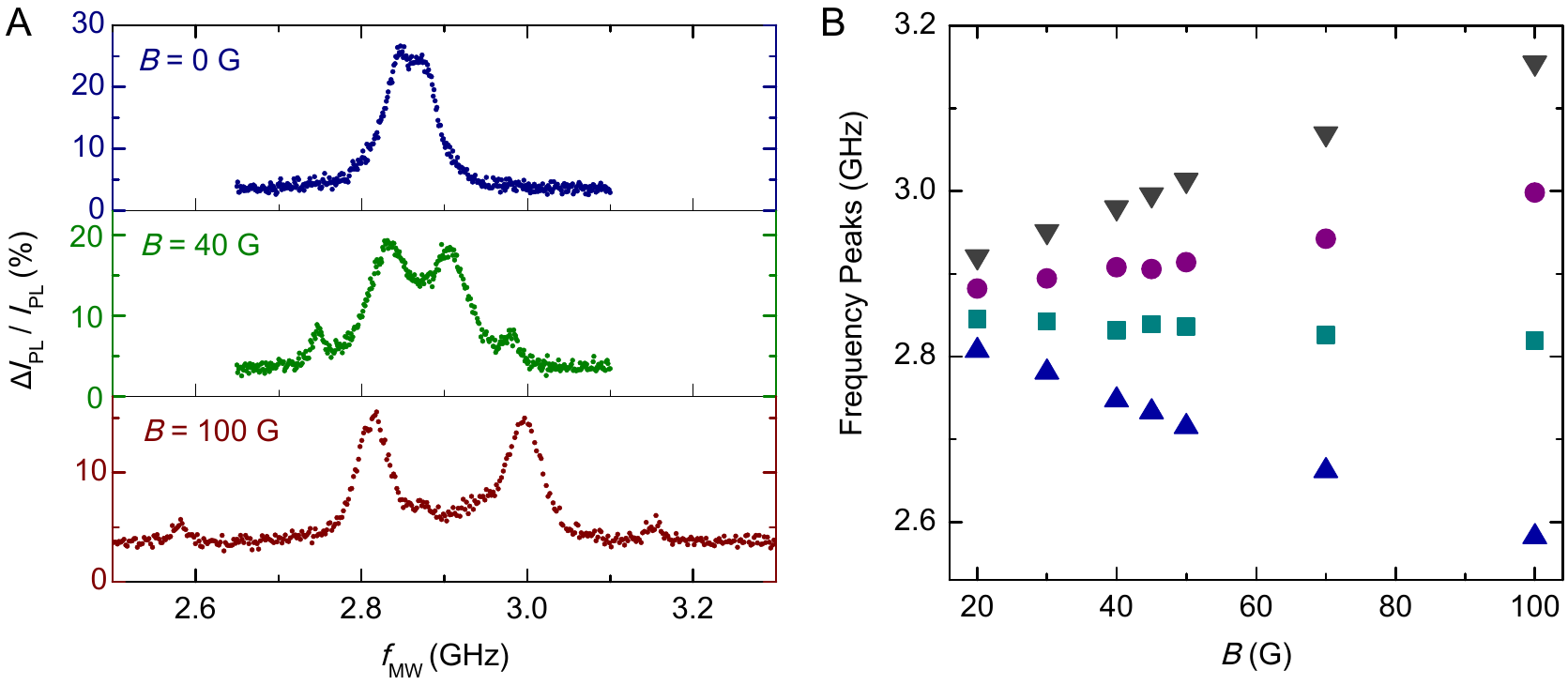}
\caption{ 
	(\textit{A})~Optically detected ESR of dry nanodiamonds drop-cast and dried on a coverslip near a microwave antenna. These nanodiamonds are not irradiated. While a single spin will split into two peaks, here we observe at least four peaks, which indicate we are measuring multiple NV centers at different orientation with respect to the applied magnetic field.  The frequency peaks of these ESR spectra are plotted in (\textit{B}) as they vary with magnetic field.
} \label{fig:Zeeman}
\end{figure*}

\section{Measurements of dry, fixed nanodiamonds}

In addition to measuring optically trapped nanodiamonds, we have also measured nanodiamonds stuck to the coverslip.  Figure~\ref{fig:Zeeman} shows  ESR of nanodiamonds that are drop-cast onto the coverslip. To mitigate the influence of non-NV-based fluorescence, we photobleach the sample with the 532~nm laser, intending to let only the non-photobleaching NV fluorescence signal remain.  The ESR signal splits into more than two peaks, indicating that the measurement ensemble includes multiple NV centers at different orientations.  These nanodiamonds are not irradiated.   We observe a higher signal-to-noise ratio for ESR spectra taken with stuck nanodiamonds than with optically trapped nanodiamonds.  This suggests that trapping dynamics and a decreased optical collection efficiency contribute to the lower signal-to-noise ratio in optically trapped nanodiamonds.

\section{Modeling the ESR spectrum}

The NV centers contained in the nanodiamond ensemble are not expected to be aligned or oriented in any particular direction.  In addition, the particles may rotate in the optical trap.   Therefore, we model the ESR spectrum by assuming the NV centers are randomly oriented. We begin by calculating the angular dependence of the excitation (\textit{i.e.} absorption) and emission of a single NV center, then calculate the ESR peak frequencies as a function of magnetic field strength and NV center orientations.  Next, we integrate the contributions of an isotropic density of NV center angles to obtain the predicted ESR spectra for a large ensemble of randomly oriented NV centers in a magnetic field aligned with the axis of the microscope objective.  Finally, we use a Markov Chain Monte Carlo approach to fit the modeled curve to the data and to extract parameters such as the estimated magnetic field and the magnetic sensitivity.

\subsection{Angular dependence of the absorption of a single NV center}
The absorption of a single transition dipole is proportional to $ | \mathbf{p} \cdot \mathbf{E} |^2$, where $\mathbf{E}$ is the electric field vector of the exciting laser beam and $\mathbf{p}$ is the dipole, which we treat classically. 
An NV center has two transition dipoles, each perpendicular to the axis of the NV center~\cite{Epstein2005}, so the combined absorption is
\[
  \text{Absorption} \propto  |\mathbf{p}_1 \cdot \mathbf{E}  |^2 +   |  \mathbf{p}_2 \cdot \mathbf{E}|^2.
\]
Although a highly focused Gaussian beam includes electric fields at multiple angles, we approximate that the electric field that excites the NV center is perpendicular to the axis of the microscope objective.  We also assume that the electric field is linearly polarized, $\mathbf{E}_\text{laser} = E_x \hat{x}$, though this assumption is merely for convenience and will not affect the calculation once the absorption is integrated over all possible NV angles. The laser polarization would be important if the magnetic field were applied along a different axis with respect to the microscope objective, but our geometry has a symmetry so the polarization of the laser is not important.  Laser polarization control could provide a route to vector magnetometry using an ensemble of randomly oriented NV centers.

For a given NV center with an angle $\theta$ with respect to the axis of the microscope objective,   we can specify with no loss of generality that  its dipole $\mathbf{p}_1$ is perpendicular to the axis of the microscope objective. Then $\mathbf{p}_2$  must be perpendicular to $\mathbf{p}_1$.  It is convenient to define these vector directions using a cross product:
\[
          \frac{\mathbf{p}_1}{ |\mathbf{p}_1 | } = 
                       \frac{ \hat{N} \times \hat{z} }
                           {| \hat{N} \times \hat{z}|} 
\quad \text{and} \quad
            \frac{\mathbf{p}_2}{ |\mathbf{p}_2|} = 
                    \frac{\hat{N} \times {\mathbf{p}_1 } }
                          {\left| {\hat{N} \times {\mathbf{p}_1 } }\right|},
\]
 where $\hat{z}$ points along the axis of the microscope objective and
$\hat{N}$ is a unit vector pointing along the direction of the
symmetry axis of the NV center,
\[
\hat{N} =  \sin \theta \cos \phi \, \hat{x} + \sin \theta \sin \phi \, \hat{y} + \cos \theta \, \hat{z},
\]
 where $\theta$ is the polar angle and $\phi$ is the azimuthal angle.  By symmetry, $|\mathbf{p}_1| = |\mathbf{p}_2|$. Then the angular dependence of the absorption of a single NV center is $1 - \cos^2 \phi \sin^2 \theta$.    Integrating over all~$\phi$, we obtain 
\begin{equation} \label{eq:absorption}
\text{Absorption} \propto 1 + \cos^2 \theta.
\end{equation}
We will integrate over all angles~$\theta$ at a later point in the calculation.

\subsection{The emission collected from a single NV center}
In order to calculate the collected emission of the transition dipole, we begin by calculating the angular part of the emission function of a dipole $\mathbf{p}$, which we treat classically.
The component of the dipole that is orthogonal to the direction vector $\hat{r}$ is 
\[
     \mathbf{p}_{\perp} = \mathbf{p} - \hat{r} \, (\hat{r}\cdot \mathbf{p}),
\]
 where 
\[
     \hat{r} = \sin \vartheta \cos\varphi \, \hat{x} + \sin \vartheta \sin \varphi \, \hat{y} + \cos \vartheta \, \hat{z}.
\]
The power radiated by the dipole is proportional to $|\mathbf{p}_{\perp} |^2$.  We integrate over the collection cone of the microscope objective to find the angular part of the power collected, 
\[
     P_1 \propto \int_0^{\vartheta_{\text{max}}} \int_0^{2 \pi}  |\mathbf{p}_{\perp} |^2 \,  \sin \vartheta \, d\varphi d\vartheta 
\]
where $P_1$ is the collected emission of one dipole and $\vartheta_{\text{max}}$ is the maximum angle at which the objective can collect light; since  $\text{NA} = n \sin {\vartheta_{\text{max}}} $ with $n = 1.515$ and $\text{NA} = 1.3$, we have ${\vartheta_{\text{max}} = 59.1^{\circ}}$.  We assume that the objective has uniform efficiency for collecting light from all angles in its collection cone. An objective with a lower NA would have a stronger angular dependence of the collection.
The result is 
\[
P_1 \propto 2.43 \, (p_x^2 + p_y^2) \, + \,  1.25 \, p_z^2,
\]
where $p_x$, $p_y$, and $p_z$ are the components of $\mathbf{p}$.
   If we account for the geometry of the two transition dipoles of the NV center, and let $\theta$ be the angle between the NV center and the axis of the microscope objective, then the collected radiation of the dipoles $\mathbf{p}_1$ and $\mathbf{p}_2$ depends on the orientation of the NV center according to 
\begin{equation} \label{eq:emission}
          \text{Collected emission} \propto  2.43 + 2.43 \cos^2 \theta + 1.25 \sin^2 \theta.
\end{equation}

\subsection{Zeeman splitting}

\begin{figure}  \centering
\includegraphics{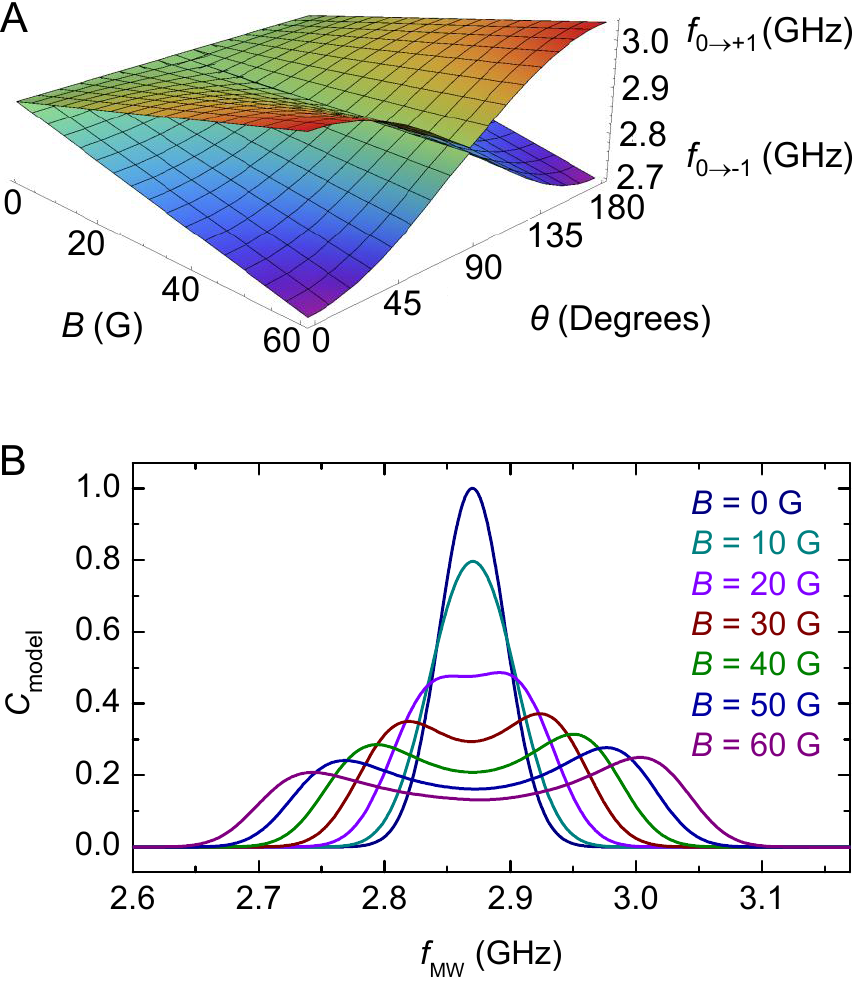}
\caption{
	(\textit{A})~The spin sublevel transition frequencies  $f_{0 \rightarrow -1}$ and $ f_{0 \rightarrow +1}$ depend on both the magnitude $B$ of the magnetic field and the angle $\theta$ between the NV symmetry axis and the magnetic field vector.  Here we assume the zero-field splitting is $D = 2.87$~GHz.
(\textit{B})~Modeled ESR spectra for an ensemble of randomly oriented NV centers, calculated from Eq.~\eqref{eq:model}, and plotted with a zero-field width of 61.7 MHz and $D$ = 2.87 GHz.  These curves are also plotted as a colormap in Fig.~{4}\textit{C}. 
} \label{fig:Modeled_peaks_spectra}
\end{figure}

The energy levels of the spin states of the ground state of the NV center are calculated directly from the ground state Hamiltonian, 
\[
	\hat{H}_\text{NV} = D \hat{S}_z^2 + g \muB \mathbf{B} \cdot \hat{\mathbf{S}},
\]
where $D = h \cdot 2.87$~GHz, $g = 2.00$,  $\muB$ is the Bohr magneton, and the components of $\hat{\mathbf{S}}$ are the spin 1 matrices. Terms of the Hamiltonian not relevant to this calculation have been suppressed. 
  The difference between spin  levels gives the frequency of the peaks measured in the ESR spectrum of a single NV center.  That is, for a given field $\mathbf{B}$, the spectrum will have peaks corresponding to
\begin{eqnarray*}
 h f_{0 \rightarrow +1}\!&=&\!E_{m_s = +1} -  E_{m_s = 0} \quad \text{and}\\
 h f_{0 \rightarrow -1}\!&=&\!E_{m_s = -1} -  E_{m_s = 0},
\end{eqnarray*}
where $E_{m_s = 0}$, $E_{m_s = -1}$, and $E_{m_s = +1}$ are the three eigenvalues of $\hat{H}_\text{NV}$.  Figure~\ref{fig:Modeled_peaks_spectra}\textit{A} shows how these spin sublevel frequencies depend on the angle of the NV center to the magnetic field.  For zero field, $E_{m_s = -1} = E_{m_s = +1}$, so the peaks are degenerate, $h f_{0 \rightarrow -1} = h f_{0 \rightarrow +1} = D$.   The frequencies split with magnetic field.  For NV centers aligned with the magnetic field, and for fields below 1000~G, the frequencies are  linear in the magnetic field: 
$ h f_{0 \rightarrow \pm 1}(\theta\!=\!0) \, = \, D \pm {g \muB}  B$, where  ${g \muB}/{h} = 2.80~\text{MHz/G}$ and $\theta$ is the angle between the NV axis and the magnetic field vector.
However, the frequency $f_{0\rightarrow-1}$ varies more with $\theta$ than $f_{0\rightarrow+1}$ does, causing an asymmetry at nonzero fields.

\subsection{ESR spectrum of a single NV center}
We assume that each NV center in the distribution has an ESR spectrum with two Gaussian functions, one peak centered at   $f_{0 \rightarrow -1}$ and the other at $f_{0 \rightarrow +1}$.  These frequencies depend on the magnetic field strength and the angle $\theta$ between the magnet and the axis of the NV center.  Note that for our geometry, where the magnet and the objective share an axis (see Fig.~1\textit{A} or \ref{fig:apparatus}), this is the same angle as the angle between the NV center and the axis of the microscope objective; for the purposes of the calculation, we assume no misalignment between the magnet and the microscope objective. The amplitude of this double-Gaussian single-NV ESR spectrum depends on the angle $\theta$ between the axis of the NV center and the axis of the microscope objective:
\[
         A_1(\theta) \propto  (1 + \cos^2 \theta) \; (2.43 + 2.43 \cos^2 \theta + 1.25 \sin^2 \theta),
\]
where Eqs.~\eqref{eq:absorption} and \eqref{eq:emission} give the angular dependence of the absorption and collected emission of the NV center.  Note that the NV center can emit a photon via either dipole, regardless of the dipole that absorbed a photon. We approximate that the microwave power affects the NV centers uniformly.  Therefore the ESR spectrum $C_1(B,\theta; \fMW)$ of a single NV has angular dependence
\[
           C_1(B,\theta;\fMW) = A_1(\theta)
                              \;  [G(f_{0 \rightarrow +1}; \fMW) + G(f_{0 \rightarrow -1} ; \fMW)],
\]
where $G(x_0;x)$ is a Gaussian function of $x$ centered at $x_0$, and $f_{0 \rightarrow +1}$ and $f_{0 \rightarrow -1}$ are functions of $B$ and $\theta$ as shown in Fig.~\ref{fig:Modeled_peaks_spectra}\textit{A}.  The widths of the two Gaussian functions must be determined empirically and are assumed to be equal to each other.  

\subsection{ESR spectrum of an isotropic ensemble of NV centers}
 To obtain the ESR spectrum of an ensemble of NV centers, $C_\text{model}(B, \fMW) $, we integrate over all angles~$\theta$,
\begin{equation} \label{eq:model}
       C_\text{model}(B, \fMW) = \int_0^\pi C_1(B,\theta; \fMW) \, \sin \theta \, d\theta,
\end{equation}
and the result is plotted in Fig.~\ref{fig:Modeled_peaks_spectra}\textit{B}.
Note that for an isotropic distribution of NV centers, more NV centers will be perpendicular to  the axis of the magnet/objective than parallel to this axis, with a probability distribution given by $\sin \theta$.  The model predicts two peaks separating and broadening as the magnetic field is increased.   The right peak is predicted to be taller and narrower than the left peak due to the asymmetry between $f_{0 \rightarrow -1}$ and $f_{0 \rightarrow +1}$.   

\subsection{Fitting the ESR curves}
To compute marginal posterior densities for the sensitivities and infer the magnetic field experienced by the NV centers in the main text, we employ the MT-DREAM$_{\text{ZS}}$ algorithm written in MATLAB
\cite{VrugtSIAM,Laloy2012}, a Markov Chain Monte Carlo approach that uses adaptive proposal distribution tuning, multiple-try sampling, sampling from the past, and snooker update on parallel chains to rapidly explore high-dimensional posterior distributions. In Markov Chain Monte Carlo, each of the $N$ chains executes a random walk through the parameter space following a modified Metropolis-Hastings rule to control whether a proposed $d$-dimensional move is accepted or rejected. Because the algorithm is ergodic and maintains detailed balance at each step, the target distribution after a burn-in period is the desired posterior probability distribution for the experiment. We find good results using the recommended settings along with $N = 6$ parallel chains with multiple-try parameter $k = 3$. Although the dimensionality of the problem ($d = 6$) is low, in practice multiple-try sampling is advantageous for faster convergence and better autocorrelation properties of the sampler output. Because the model relies on the numerical convolution over the orientation angle of the NV centers, the evaluation of the posterior probability density and the estimates of its derivative can be slow to compute. We vectorized the computation of both the Hamiltonian eigenvalues and numerical integrations over $\theta$ for each modeled frequency curve and used an NVIDIA GTX-440 graphics processing unit along with MATLAB software package \verb"Jacket" from Accelereyes to greatly increase the speed of computations of the posterior density. Convergence to the target distribution was assessed both graphically and with the Gelman-Rubin statistic, $\hat{R} < 1.02$ \cite{GelmanRubin1992}. The point estimates for the magnetic field~($B$), the homoscedastic normal error at each datapoint~($\sigma$), and other parameters are computed from the respective sample empirical means, and the highest probability density intervals are computed using the method of Chen and Shao~\cite{ChenShao1999}.

\begin{figure} \centering
\includegraphics{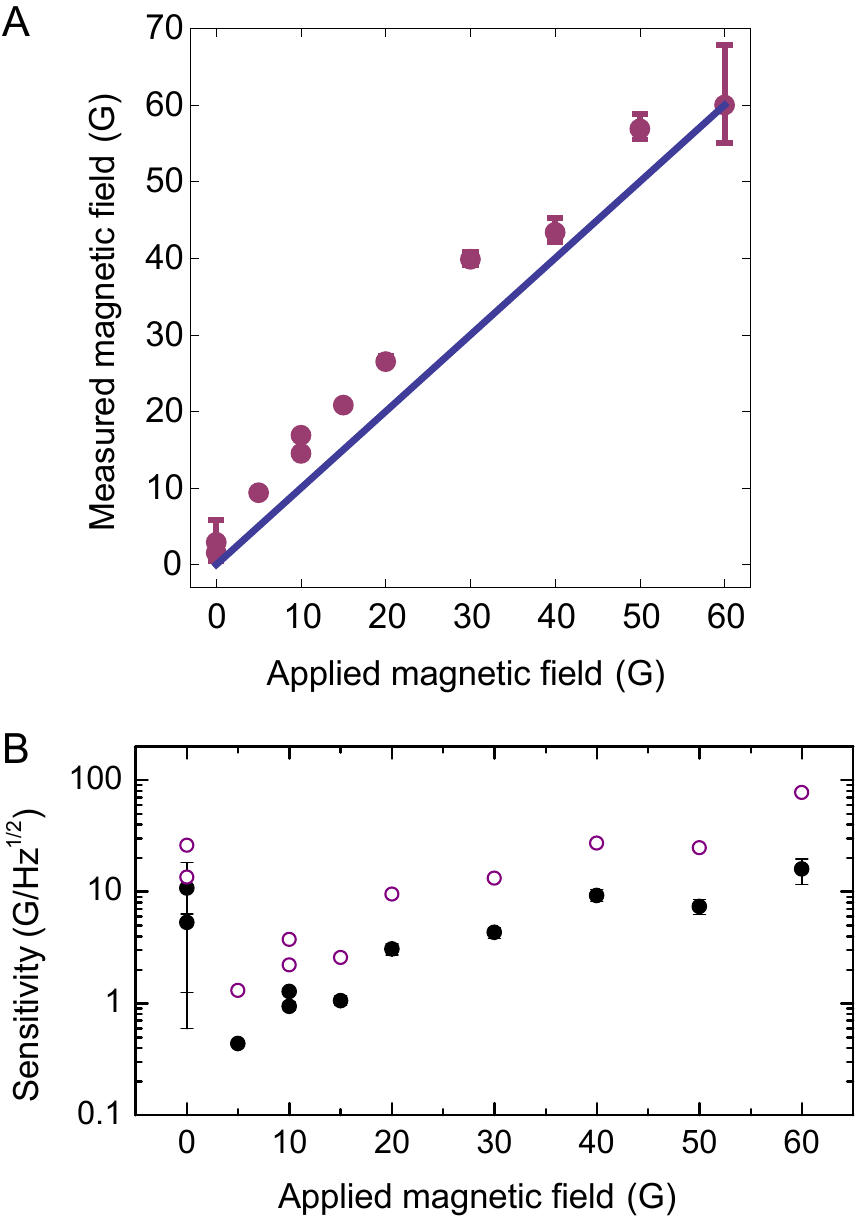}
\caption{
(\textit{A})~Magnetic field measured by the trapped NV ensemble versus applied magnetic field. Plotted error bars are 95\% highest probability density intervals. The measured values are inferred from the model fitting while the applied values are set by an \textit{ex situ} calibration. The solid line of unit slope and zero offset is used to compare the data against the ideal measurement.
(\textit{B})~The optimum sensitivity (black circles with error bars) and the demonstrated sensitivity (open purple circles) of the optically trapped nanodiamond-ensemble magnetometer.
} \label{fig:Bfield_Sensitivity}
\end{figure}

The mean of the marginal posterior density of $B$ is plotted in Fig.~\ref{fig:Bfield_Sensitivity}\textit{A}. The 95\% highest probability density intervals are plotted as error bars. The plot demonstrates the ability of the apparatus to sense the applied calibrated magnetic field. The discrepancy between the applied field and measured field is about $\sim$5\,{G}, and appears to be a repeatable, constant offset. We attribute this error to the magnetic piezostage used in the experiment, whose field is not accounted for in the calibration. Another explanation may be some deficiency in the model, but the linearity of the sensed magnetic field versus applied magnetic field seems to discount this as the primary issue.

\section{Magnetic sensitivity}

For theoretical sensitivity calculations, we consider the optimal measuring frequency to be the frequency at which the derivative $\frac{\partial C}{\partial B}$ has maximum magnitude. One can imagine constructing a measurement scheme for magnetometry that occurs at this single frequency to detect small changes in the applied magnetic field $B$. Having inferred the noise in the contrast of our measurement from the analysis, we use the $1\sigma$ change as the critical value for the minimum detectable change in magnetic field. Given the value of $\sigma$ inferred from our experiments as a function of field, we calculate a theoretical sensitivity for such an idealized measurement (black circles in Fig.~\ref{fig:Bfield_Sensitivity}\textit{B}. The computation proceeds by taking individual samples from the Markov Chain Monte Carlo output and computing the maximum magnitude of the partial derivative with respect to $B$ at each parameter space sample, and dividing the corresponding sample of $\sigma$ by this numerical derivative to generate the posterior distribution of the sensitivity $\eta$. The heavy tails of $\eta$ at zero field, which arise from the vanishing of the numerical derivative $\frac{\partial C}{\partial B}$ as $B$ tends to zero, explain the large error bars found there. As a practical check, we can estimate the achieved sensitivity of the magnetometer on the basis of a scheme of taking ESR sweeps in the same fashion taken in the main text. By computing the fitting uncertainty from the marginal distribution of $B$ and scaling it by the square root of the acquisition time, we can calculate an empirical measure of the instrument sensitivity, i.e. $\eta_\text{empirical} = \sigma_{B} \sqrt{\Delta t}$. These calculated values are additionally plotted in Fig.~\ref{fig:Bfield_Sensitivity}\textit{B} (open purple circles) for comparison to the theoretical values. In the empirical scheme, the microwave frequency is swept across the spin resonances but also measures the off-resonant signal that contributes almost no information to the determination of $B$. Thus, the empirical measure is necessarily less efficient in its use of the resource of acquisition time, with a commensurately worse sensitivity. The observation that the theoretical sensitivities proposed in the manuscript are only a few times better than the empirical sensitivities demonstrated directly from the fitting ensures that the theoretical estimates are not unreasonable.


\begin{thebibliography}{10}

\bibitem{Jelezko2004}
Jelezko F, Gaebel T, Popa I, Gruber A, Wrachtrup J (2004) Observation of
  coherent oscillations in a single electron spin. \emph{Physical Review
  Letters} 92:76401.

\bibitem{Degen2008}
Degen CL (2008) Scanning magnetic field microscope with a diamond single-spin
  sensor. \emph{Applied Physics Letters} 92:243111.

\bibitem{Taylor2008}
Taylor JM, {et~al.} (2008) High-sensitivity diamond magnetometer with
  nanoscale resolution. \emph{Nature Physics} 4:810--816.

\bibitem{Balasubramanian2008}
Balasubramanian G, {et~al.} (2008) Nanoscale imaging magnetometry with
  diamond spins under ambient conditions. \emph{Nature} 455:648--651.

\bibitem{Maze2008}
Maze JR, {et~al.} (2008) Nanoscale magnetic sensing with an individual
  electronic spin in diamond. \emph{Nature} 455:644--647.

\bibitem{Maertz2010}
Maertz BJ, Wijnheijmer AP, Fuchs GD, Nowakowski ME, Awschalom DD (2010) Vector
  magnetic field microscopy using nitrogen vacancy centers in diamond.
  \emph{Applied Physics Letters} 96:092504.

\bibitem{Schoenfeld2011}
Schoenfeld RS, Harneit W (2011) Real time magnetic field sensing and imaging
  using a single spin in diamond. \emph{Physical Review Letters} 106:030802.

\bibitem{Pham2011}
Pham LM, {et~al.} (2011) Magnetic field imaging with nitrogen-vacancy
  ensembles. \emph{New Journal of Physics} 13:045021.

\bibitem{Arcizet2011}
Arcizet O, {et~al.} (2011) A single nitrogen-vacancy defect coupled to a
  nanomechanical oscillator. \emph{Nature Physics} 7:879--883.

\bibitem{Rondin2012}
Rondin L, {et~al.} (2012) Nanoscale magnetic field mapping with a single
  spin scanning probe magnetometer. \emph{Applied Physics Letters} 100:153118.

\bibitem{Maletinsky2012}
Maletinsky P, {et~al.} (2012) A robust scanning diamond sensor for
  nanoscale imaging with single nitrogen-vacancy centres. \emph{Nature
  Nanotechnology} 7:320--324.

\bibitem{Dolde2011}
Dolde F, {et~al.} (2011) Electric-field sensing using single diamond
  spins. \emph{Nature Physics} 7:459--463.

\bibitem{Toyli2012}
Toyli DM, {et~al.} (submitted) Persistence of single spin coherence above
  600{K} in diamond. arXiv:12014420.

\bibitem{Yu2005}
Yu SJ, Kang MW, Chang HC, Chen KM, Yu YC (2005) Bright fluorescent
  nanodiamonds: No photobleaching and low cytotoxicity. \emph{Journal of the
  American Chemical Society} 127:17604--17605.

\bibitem{Fu2007}
Fu CC, {et~al.} (2007) Characterization and application of single
  fluorescent nanodiamonds as cellular biomarkers. \emph{Proceedings of the
  National Academy of Sciences} 104:727--732.

\bibitem{Liu2007}
Liu KK, Cheng CL, Chang CC, Chao JI (2007) Biocompatible and detectable
  carboxylated nanodiamond on human cell. \emph{Nanotechnology} 18:325102.

\bibitem{Schrand2007}
Schrand AM, {et~al.} (2007) Are diamond nanoparticles cytotoxic? \emph{The
  Journal of Physical Chemistry B} 111:2--7.

\bibitem{Chang2008}
Chang YR, {et~al.} (2008) Mass production and dynamic imaging of
  fluorescent nanodiamonds. \emph{Nature Nanotechnology} 3:284--288.

\bibitem{McGuinness2011}
McGuinness LP, {et~al.} (2011) Quantum measurement and orientation
  tracking of fluorescent nanodiamonds inside living cells. \emph{Nature
  Nanotechnology} 6:358--363.

\bibitem{vanderSar2009}
van~der Sar T, {et~al.} (2009) Nanopositioning of a diamond nanocrystal
  containing a single nitrogen-vacancy defect center. \emph{Applied Physics
  Letters} 94:173104.

\bibitem{Barth2009}
Barth M, N\"{u}sse N, L\"{o}chel B, Benson O (2009) Controlled coupling of a
  single-diamond nanocrystal to a photonic crystal cavity. \emph{Optics
  Letters} 34:1108--1110.

\bibitem{Cuche2009Near}
Cuche A, {et~al.} (2009) Near-field optical microscopy with a
  nanodiamond-based single-photon tip. \emph{Optics Express} 17:19969--19980.

\bibitem{Neuman2004}
Neuman KC, Block SM (2004) Optical trapping. \emph{Review of Scientific
  Instruments} 75:2787--2809.

\bibitem{Sun2001}
Sun CK, Huang YC, Cheng PC, Liu HC, Lin BL (2001) Cell manipulation by use of
  diamond microparticles as handles of optical tweezers. \emph{Journal of the
  Optical Society of America B-Optical Physics} 18:1483--1489.

\bibitem{Block1990}
Block S, Goldstein L, Schnapp B (1990) Bead movement by single kinesin
  molecules studied with optical tweezers. \emph{Nature} 348:348--352.

\bibitem{Ericsson2000}
Ericsson M, Hanstorp D, Hagberg P, Enger J, Nystr{\"o}m T (2000) Sorting out
  bacterial viability with optical tweezers. \emph{Journal of Bacteriology}
  182:5551--5555.

\bibitem{Abbondanzieri2005}
Abbondanzieri E, Greenleaf W, Shaevitz J, Landick R, Block S (2005) Direct
  observation of base-pair stepping by {RNA} polymerase. \emph{Nature}
  438:460--465.

\bibitem{Svoboda1993}
Svoboda K, Schmidt C, Schnapp B, Block S, {et~al.} (1993) Direct
  observation of kinesin stepping by optical trapping interferometry.
  \emph{Nature} 365:721--727.

\bibitem{Gruber1997}
Gruber A, {et~al.} (1997) Scanning confocal optical microscopy and
  magnetic resonance on single defect centers. \emph{Science} 276:2012--2014.

\bibitem{Manson2006}
Manson NB, Harrison JP, Sellars MJ (2006) Nitrogen-vacancy center in diamond:
  Model of the electronic structure and associated dynamics. \emph{Physical
  Review B} 74:104303.

\bibitem{Bradac2010}
Bradac C, {et~al.} (2010) Observation and control of blinking
  nitrogen-vacancy centres in discrete nanodiamonds. \emph{Nature
  Nanotechnology} 5:345--349.

\bibitem{Braun2002}
Braun D, Libchaber A (2002) Computer-based photon-counting lock-in for phase
  detection at the shot-noise limit. \emph{Optics Letters} 27:1418--1420.

\bibitem{Arecchi1966}
Arecchi F, Gatti E, Sona A (1966) Time distribution of photons from coherent
  and {G}aussian sources. \emph{Physics Letters} 20:27--29.

\bibitem{Laloy2012}
Laloy E, Vrugt JA (2012) High-dimensional posterior exploration of hydrologic
  models using multiple-try {DREAM}$_{\text{(zs)}}$ and high-performance
  computing. \emph{Water Resources Research} 48:W01526.

\bibitem{Shelton2005}
Shelton WA, Bonin KD, Walker TG (2005) Nonlinear motion of optically torqued
  nanorods. \emph{Physical Review E} 71:036204.

\bibitem{Duan2012}
Duan X, {et~al.} (2012) Intracellular recordings of action potentials by
  an extracellular nanoscale field-effect transistor. \emph{Nature
  Nanotechnology} 7:174--179.

\bibitem{Brites2012}
Brites CDS, {et~al.} (2012) Thermometry at the nanoscale. \emph{Nanoscale}.

\bibitem{Hall2012}
Hall LT, {et~al.} (2012) High spatial and temporal resolution wide-field
  imaging of neuron activity using quantum {NV}-diamond. \emph{Nature
  Scientific Reports} 2:401.

\bibitem{Epstein2005}
Epstein RJ, Mendoza FM, Kato YK, Awschalom DD (2005) Anisotropic interactions
  of a single spin and dark-spin spectroscopy in diamond. \emph{Nature Physics}
  1:94--98.

\bibitem{Dreau2011}
Dr\'eau A, {et~al.} (2011) Avoiding power broadening in optically detected
  magnetic resonance of single {NV} defects for enhanced dc magnetic field
  sensitivity. \emph{Physical Review B} 84:195204.

\bibitem{ChenShao1999}
Chen MH, Shao QM (1999) Monte {C}arlo estimation of {B}ayesian credible and
  {HPD} intervals. \emph{Journal of Computational and Graphical Statistics}
  8:69--92.

\bibitem{VrugtSIAM}
Vrugt JA, Laloy E, ter Braak CJF, Hyman JM (2012) Posterior exploration using
  differential evolution adaptive {M}etropolis with sampling from past states.
  To be submitted to \textit{SIAM Optimization}.

\bibitem{GelmanRubin1992}
Gelman A, Rubin DB (1992) Inference from iterative simulation using multiple
  sequences. \emph{Statistical Science} 7:457--472.

\end{thebibliography}
\end{document}